\documentclass[aps, pre,amsmath,amssymb,floatfix]{revtex4}

\newif\ifpdf\ifx\pdfoutput\undefined\pdffalse\else\pdfoutput=1\pdftrue\fi

\renewcommand{\vec}[1]{{\bf #1}}
\newcounter{abc}
\renewcommand{\theequation}{\arabic{equation}\alph{abc}}

\newcommand{\rset}{\{\vec{r}\}}

\newcommand{\etaset}{\{\eta\}}

\newcommand{\uset}{\{\vec{u}\}}
\newcommand{\usetstar}{\{\vec{u}\}_{\star}}

\newcommand{\usetstarprime}{\{\vec{u}\}^{\prime}_{\star}}

\newcommand{\RsetCS}{\{ \vec{R}\}^{CS}}

\newcommand{\RsetF}{\{ \vec{R}\}^{F}}

\newcommand{\Rsetgamma}{\{ \vec{R}\}^{\gamma}}
\newcommand{\Rsetgammaprime}{\{ \vec{R}\}^{\gamma^{\prime}}}

\newcommand{\zcaltilde}[4]{{\tilde{\cal{Z}}}(#1,#2,#3,#4)}
\newcommand{\zcal}[4]{{\cal{Z}}_{#1}(#2, #3,#4)}
\newcommand{\zsimp}[4]{{{Z}}_{#1} (#2, #3, #4)}
\newcommand{\zratio}{{\mathcal{R}}_{\mbox{\sc {f,cs}}}}

\newcommand{\equalnorm}{\propto}

\usepackage{graphicx}
\usepackage{bm}
\usepackage{epsfig}

\newcommand{\be}{\begin{equation}}

\newcommand{\ee}{\end{equation}}

\begin{document}

\title{\bf Freezing line of the Lennard-Jones fluid: a Phase Switch Monte Carlo study}

\author{Graham C. McNeil-Watson} 
\author{Nigel B. Wilding}
\affiliation{Department of Physics, University of Bath, Bath BA2 7AY, United Kingdom}

\date{\today}

\begin{abstract} 

We report a Phase Switch Monte Carlo (PSMC) method study of the
freezing line of the Lennard-Jones (LJ) fluid. Our work generalizes to
soft potentials the original application of the method to hard sphere
freezing, and builds on a previous PSMC study of the LJ system by
Errington (J. Chem. Phys. {\bf 120}, 3130 (2004)). The latter work is
extended by tracing a large section of the Lennard-Jones freezing
curve, the results for which we compare to a previous Gibbs-Duhem
integration study. Additionally we provide new background regarding the
statistical mechanical basis of the PSMC method and extensive
implementation details.


\end{abstract} 
\maketitle
\setcounter{totalnumber}{10}

\section{Introduction and background}

\label{sec:intro}

The freezing of a disordered fluid into an ordered crystalline solid
undoubtedly represents one of the most spectacular examples of
thermodynamics in action.  But despite its ubiquity and familiarity,
key aspects of the phenomenon remain poorly understood across a variety
of systems \cite{MONSON00}. Principal among the reasons for this is the
difficulty of obtaining accurate simulation estimates for fluid-solid
phase coexistence properties. In this section we outline the most
commonly used contemporary approaches for tracing freezing boundaries,
identify their key strengths and weaknesses, and highlight recent
developments in the search for improvements.

The staple method for locating fluid-solid coexistence is thermodynamic
integration (TI) \cite{FRENKEL84,FRENKELSMIT,MONSON00}. Here, the free
energy of each phase (fluid (F) and crystalline solid (CS)) is computed
for states covering a range of densities, using integration methods
which connect their thermodynamic properties with those of reference
states of known free energy. The two branches of the free-energy are
then matched to determine the freezing parameters. While there is much
to commend TI (principally its simplicity) it turns out to be less than
ideal in a number of respects \cite{OTHERAPPROACHES}. These have been
documented in detail elsewhere (see eg. \cite{BRUCE03}); but briefly,
the main difficulty is one of identifying a suitable reference state
together with a reversible phase space pathway by which it can be
reached; a poor choice of integration path may encounter singularities
 --both real and artificial \cite{singularities}-- which compromise the
calculation.  Additionally, corrections may be required to allow for
the fact that the path does not quite reach the idealized reference
state \cite{FRENKEL84,polson}. Moreover, the method focuses on the
absolute free energies of the phases \cite{BARROSO02}, rather than the
quantity of interest --their free energy {\em difference}. Finally, there
currently exists no reliable and comprehensive method for estimating
errors on phase boundaries computed via TI.

Obtaining a whole F-CS phase boundary using TI is a computationally
challenging task. However, knowledge of one point on the F-CS
coexistence boundary (obtained, for instance, by using TI) can be used to
bootstrap a Gibbs-Duhem integration (GDI) scheme to trace the entire
curve without further calculation of free energies. The GDI method
\cite{KOFKE93A,KOFKE93B,AGRAWAL95} exploits the generalized
Clausius-Clapeyron equation to express the slope of the phase boundary
entirely in terms of single-phase averages. This is clearly a virtue
since it avoids the need for two-phase sampling. However, {\em without}
any `reconnection' of the two configuration-spaces at subsequent
simulation state points, the GDI approach offers no feedback on
integration errors. Since there will generally exist a {\em band} of
metastable states on each side of the phase boundary, it is possible
for the integration to wander significantly from the true boundary with
no indication that anything is wrong. As a result, the calculation of
meaningful uncertainties is problematic. GDI has been used in a number
of studies, most notably in the context of freezing of hard and soft
spheres \cite{GDIexamples}.

More recently, attention has shifted to developing methods for tackling
the problem of locating F-CS coexistence via {\em direct} measurements
of free energy differences. To do so, one must construct a reversible
sampling path between the phases. The obvious ``physical'' path,
traversing the region where both phases coexist --whilst practicable
when the two phases share the same symmetry (see eg.
ref.~\cite{WILDING95}) --turns out to be computationally problematic in
the F-CS context. The main reasons for this are the large degree of
metastability of the two phases, the extended timescale for the
crystallization process, and the tendency of any crystal formed to
exhibit vacancies, interstitials and other defects. Accordingly, recent
effort has focussed on trying to identify alternative inter-phase
routes for the F-CS problem.

One such alternative route, called constrained fluid-$\lambda$
integration has been proposed by Grochla\cite{GROCHLA04} and extended
by Elke et al \cite{ELKE05}. It involves the controlled transformation
of the fluid phase to the solid phase in a series of stages. During
the initial stage the fluid is gradually changed to a weakly
interacting fluid by reducing the strength of interparticle
interactions. At the next stage, a set of Gaussian potential wells of
prescribed width are switched on at the sites of a crystalline lattice
of the appropriate symmetry. Simultaneously, the volume of the system
is gradually changed from a value typical of the fluid to one typical
of the solid. The particles of the weakly interacting fluid are then
allowed to diffuse to the lattice sites, where they are captured by the
Gaussian wells. The final stage involves gradually restoring the
particle interactions to full strength, whilst simultaneously switching
off the Gaussian potential wells. Integration of free energy
derivatives is used to estimate the free energy difference along the
path. 

The constrained fluid-$\lambda$ integration method was tested by
application to the Lennard-Jones fluid where it was used to determine
two coexistence state points. This test led the author to conclude that
the method ``cannot be said to be computationally efficient''
\cite{GROCHLA04}, at least in its present form. This finding presumably
reflects the rather long and fragmented nature of the inter-phase path.
Although each of the stages of the path was initially reported to be
reversible \cite{GROCHLA05,ELKE05}, further application of the method
\cite{GROCHLA05} found this to be contingent on the correct tuning (via
trial and error) of four separate constants relating to the
characteristics of the Gaussian potential wells. Indeed, it is not
clear to us that one can generally expect {\em a-priori} that the
stated path will be trouble-free; for example, it seems hard to rule
out that the transformation of the original fluid to a weakly
interacting fluid might intersect a first order (pseudo) liquid-vapor
transition. Furthermore we speculate that for large systems the
reliance on particle diffusion to translate particles to unoccupied
lattice sites may become problematic.

Very recently, Mastny and de Pablo \cite{MASTNY05} have proposed a
method based on measurements of the density of states, which aims to
directly estimate the free energy difference of F and CS phases. The
rationale for the method is its authors' assertion that: ``To connect
the free energy of the solid and liquid phases, there must exist a
portion of energy and volume space that can be simultaneously sampled
by both solid and liquid phases''. To exploit this premise, a series of
simulations of the Lennard-Jones system were performed, each restricted
to --and sampling the configuration space of-- one of a set of
overlapping ``windows'' in the space of the total energy $E$ and volume
$V$. Successive windows were positioned appropriately such as to form a
path linking the two well-separated regions of $E,V$ associated with
the typical configurations of the respective phases.  Along this path,
a central range of $E,V$ was indeed reported to be found for which both
liquid and solid phase configurations could be sampled. A simple
average \cite{TEMPERINGSWAPS} of the solid phase and liquid phase
density of states was accumulated in the central range and joined for
continuity to the measured forms of the liquid and solid density of
states on either side. 

In our opinion, the existence of a range of $E$ and $V$ that can be
sampled by both liquid and solid phases is insufficient to connect the
respective branches of the density of states because these are (like
their underlying sets of characteristic configurations), fundamentally
{\em distinct}. Instead, for a proper connection, the phases must be
linked via a continuous (and reversible) path through configuration
space. In the context of the method of Mastny and de Pablo, this
necessitates repeated (back and forth) {\em transitions} between the two
pure phases. Since no mention is made in ref.~\cite{MASTNY05} of any
such transitions, the validity of the method would appear to be 
questionable \cite{MASTNYCONTACT}.

Another very recent approach to the F-CS problem, which bears some
resemblance to that of ref.~\cite{MASTNY05}, is the ``multi-NpH'' method
of Escobedo \cite{ESCOBEDO05}. Here a path is constructed in the {\em
enthalpy} of the system, spanning the range of values between those
typical of the respective equilibrium phases at some prescribed
pressure. The system temperature is ascertained at each point along the
path, and exhibits a discontinuity at some value of the enthalpy as the
favored phase of the system changes. TI with respect to the temperature
variations along the path yields the free energy difference. The
problem reported with this method was that it was apparently necessary
for equilibration purposes to initialize the system in the CS phase at
all points along the enthalpy path. However, it was difficult to be
sure whether, for a given enthalpy value, the system had relaxed to the
phase of minimum free energy, and hence whether the TI result was
unbiased. This seems to us to be a manifestation of the familiar
drawback of TI, namely the difficulty of finding a reversible
integration path.

Phase Switch Monte Carlo (PSMC) \cite{WILDING00} is a further
relatively recent method designed to directly link F and CS phases via
a reversible sampling path. The method builds on previous work
\cite{LSpapers} in which it was demonstrated that the configuration
spaces associated with two phases of a many-body system can both be
visited in a single Monte Carlo simulation, by harnessing extended
sampling methods to facilitate a direct switch from one phase to the
other. The method, which is quite generally applicable, was initially
deployed in a study of hard sphere freezing \cite{WILDING00}.
Subsequently, it was applied to soft potentials by Errington
\cite{ERRINGTON04}, who used it to calculate two points on the freezing
line of the Lennard-Jones system for a number of system sizes.

The present paper complements and extends Errington's PSMC study of the
LJ system. It is organized as follows.  In sec.~\ref{sec:statmech} we
review some of the key results in the statistical mechanical
formulation of F and CS phase configurational weights, and show how to
construct, on this basis, a theoretical framework for computational
estimation of free energy differences. Sec~\ref{sec:method} describes
how this framework is exploited in principle, and realized in practice,
by the PSMC method. An application of the method to
the Lennard-Jones system is reported in sec.~\ref{sec:results} where we
present estimates of large portions of the freezing curve for a number
of system sizes. These we compare with the results of a previous GDI
study, bootstrapped using TI at a single coexistence point. Finally
sec.~\ref{sec:concs} details our conclusions and the prospects for
further applications.

\section{Statistical Mechanics}

\label{sec:statmech}

Within the framework of statistical mechanics, the free energy of a
given phase is expressible in terms of its {\em configurational
weight}. Information regarding this weight can be accumulated via a
simulation in the course of an exploration (sampling) of the
microstates of the phase. While for fluid phases such an exploration
encounters no great obstacles, a complication arises in the case of
crystalline solids. Specifically, it transpires that the relevant phase
space of the solid is effectively fragmented into a number of mutually
inaccessible regions. Each fragment corresponds to a distinct
permutation of the particles with respect to the lattice sites. The
solid phase configurational weight can only be estimated for the single
fragment in which it is initiated. Since on symmetry grounds the weight
contribution of every fragment is identical, the overall
configurational weight of the solid is obtained by multiplying the
measured weight of one fragment by the number of fragments. In this
section we first show how to correctly count this fragment number. We
then proceed to describe how the ratio of configurational weights of
two phases is related to the ratio of their {\em a priori}
probabilities (a quantity directly accessible to simulation) and 
thence to the free energy difference.

Consider a periodic system of $N$ classical particles confined to a
volume $V$, variable under a prescribed constant external pressure $p$,
and in equilibrium with a heat bath at prescribed constant temperature
$T$. Within this constant-$NpT$ ensemble the configurational weight of a
phase may be written as

\begin{equation}
\label{eq:zcaldef}
{\cal Z}_{\gamma}(N,p,T)  = \int_{0}^{\infty} dV e^{- p V} Z_{\gamma}(N,V,T)\:.
\end{equation}
Here, $\gamma$ labels the phase, while
\begin{equation}
\label{eq:zsimpdef}
Z_{\gamma}(N,V,T)
= \prod_{i=1}^{N}\int_{V,\gamma} d \vec{r}_{i}e^{- \Phi( \{\vec{r}\})} \:,
\end{equation}
with $\Phi$ the configurational energy \cite{units}. The $\gamma$-label on the
integral stands for some {\em configurational constraint} that picks
out configurations $\{\vec{r}\}$ that `belong' to phase $\gamma$. In the
present work we shall be concerned with  phases which are either fluid
($\gamma={\rm F}$) or crystalline solid ($\gamma={\rm CS}$) in character, and choose to formulate the
constraint in a form which reflects the situation actually encountered
in Monte Carlo simulations of single phases.  Specifically, let
$\vec{R}_{1}^{\gamma} \ldots \vec{R}_{N}^{\gamma} \equiv \{
\vec{R}\}^{\gamma}$ denote some representative configuration of phase
$\gamma$. Then the constraint may be regarded as picking out those
configurations which can be reached from $\{ \vec{R}\}^{\gamma}$ on the
simulation timescale. The timescale is presumed to be sufficiently long
to explore `one phase' but still short compared to (unaided) inter-phase
traverses. 

It is convenient to use the sites defined by $\{ \vec{R}\}^{\gamma}$ as
the origins of the particle coordinates. Thus we define a set of
displacement vectors $\{\vec{u}\}$ where

\begin{equation}
\label{eq:origin}
\vec{u}_i\equiv \vec{r}_i - \vec{R}^{\gamma}_i\:,
\end{equation}
and write $\Phi{\gamma} (\{\vec{u}\} ) \equiv \Phi (\{ \vec{R}^{\gamma} + \vec{u}\} )$.
In the case of the fluid phase, particles have the run of the entire
system and hence {\em all}  contributing configurations are reachable
from any one; we may write simply

\begin{equation} 
\label{eq:zsimpF}
Z_{F}(N,V,T)
= \prod_{i=1}^{N}\int_{V,\{ \vec{R}\}_{F}} d \vec{u}_{i} e^{- \Phi_F (\{\vec{u}\})} \:,
\end{equation}
where $\{ \vec{R}\}^{F}$ is some specific but arbitrary fluid
configuration,  which can be  selected at random in the course of MC
exploration of the  fluid phase. 

For the crystalline phase it is natural to choose $\{ \vec{R}\}^{CS}$
to define the sites of a lattice of the appropriate symmetry  and
scale.  But here one must recognize that the complete CS configuration
space actually comprises a number of distinct mutually inaccessible
{\em fragments}  \cite{theyaredistinct}  corresponding essentially to
the different permutations  of particles amongst lattice sites. By
symmetry each fragment should contribute equally to the configurational
weight. But in general, the typical timescale for particle interchanges
between lattice sites greatly exceeds the accessible simulation timescale. In
these circumstances  MC simulation will visit (and thus count) only the
states within the fragment in which it is initiated. The total
configurational weight of the CS phase must therefore be obtained by
multiplying the measured contribution of one fragment by the total
number of fragments.  Since global translation (permitted by the
boundary conditions) ensures that {\em one} fragment includes {\em all}
possible locations of any chosen particle, the number of fragments is
the number of ways of assigning the other $N-1$ particles to $N-1$
Wigner-Seitz cells of some underlying notional fixed lattice
\cite{NOTNFAC}. This number is $(N-1)!$. Thus

\begin{equation}
\label{eq:zsimpCS}
Z_{CS}(N,V,T)
= (N-1)!\prod_{i=1}^{N}\int_{V,\{ \vec{R}\}^{CS}} d \vec{u}_{i} 
e^{- \Phi_{CS}(\{\vec{u}\})} \:.
\end{equation}

In order to make contact with the simulation methodology to be
described below, it is expedient to define the total {\em a priori}
probability of phase $\gamma$. This is given by the ratio of the
configurational weight of phase $\gamma$ to the total weight of the two
phases \cite{BRUCE03}:

\begin{equation}
P(\gamma|N,p,T)=\frac{{\cal Z}_{\gamma}(N,p,T)}{{\cal Z}_F(N,p,T)+{\cal Z}_{CS}(N,p,T)}\:.
\end{equation}
Combining this last equation with eqs.~\ref{eq:zcaldef}, \ref{eq:zsimpF} and
\ref{eq:zsimpCS}, allows the ratio of the configurational weights of the two phases to be expressed
in terms of the ratio of their {\em a priori} probabilities \cite{LSpapers}:
\begin{widetext}
\begin{eqnarray}
\label{eq:zrat} 
{\mathcal{R}}_{\mbox{\sc {f,cs}}} &\equiv &\frac {{\cal Z}_{F}(N,p,T)}   {{\cal Z}_{CS}(N,p,T)}=\frac{P(F| N,p,T)}{P(CS|N,p,T)} \\
&=&\frac {\int_{0}^{\infty} dV e^{- p V} 
\prod_{i=1}^{N}\int_{V,\{ \vec{R}\}^{F}} d \vec{u}_{i}
e^{- \Phi_{F}(\{\vec{u}\})} } {(N-1)!\int_{0}^{\infty} dV e^{- p V} 
\prod_{i=1}^{N}\int_{V,\{ \vec{R}\}^{CS}} d \vec{u}_{i}
e^{- \Phi_{CS}(\{\vec{u}\})} } \nonumber \:.
\end{eqnarray}
\end{widetext}

The link between the configurational weight of a phase and its Gibbs
free energy is forged by the definition:

\begin{equation}
{\cal Z}_{\gamma}(N,p,T)\equiv e^{-G_\gamma(N,p,T)}\:.
\end{equation}
Inserting this into eq.~\ref{eq:zrat}, the Gibbs
free-energy-density difference between the phases can be written in the
(strategically suggestive) form:

\begin{equation} 
\label{eq:gibbs} 
\Delta g \equiv g_{CS}(N,p, T)-g_{F} (N,p,T)  \equiv \frac{1}{N} \ln {\mathcal{R}}_{\mbox{\sc {f,cs}}}\:.
\end{equation}

Eqs.~\ref{eq:zrat} and~\ref{eq:gibbs} provide a foundation for MC
studies of fluid-solid phase coexistence. Their formal promise is as
follows: given an inter-phase sampling scheme that visits both the F
and CS phases, ${\mathcal {R}}_{\mbox{\sc {f,cs}}}$ (and hence $\Delta
g$) may be obtained as the ratio of the {\em a priori} probabilities of the
phases. The latter quantity can be estimated simply as the ratio of the
number of visits to each phase recorded in the course of a simulation. In order to exploit this framework, however,
one must first design a singularity-free phase space path linking the F and
CS phases, and then formulate a sampling strategy to traverse it. Phase
Switch Monte Carlo is one realization of such a scheme.

\section{Phase Switch Monte Carlo}

\label{sec:method}

The relative stability of the F and CS phases is determined by the
ratio of the associated configurational weights ${\mathcal
{R}}_{\mbox{\sc {f,cs}}}$ (Eq.~\ref{eq:zrat}).  To measure that ratio
one needs a MC procedure which visits both solid and fluid regions of
configuration space in the course of a single simulation run. The phase
switch MC method is a general approach that facilitates such two-phase
sampling. Its principal feature is a phase space `leap' \cite{LSpapers}
that directly maps a pure phase configuration of one phase onto a pure
phase configuration of the other. The motivation for choosing such an
inter-phase route for the F-CS problem is that it circumvents
interfacial (mixed-phase) configurations and their attendant sampling
difficulties (see sec.~\ref{sec:intro}). In this section we describe
how to apply the phase switch method in this context.
  
\begin{figure}[h]
\includegraphics[type=eps,ext=.eps,read=.eps,width=9.0cm,clip=true]{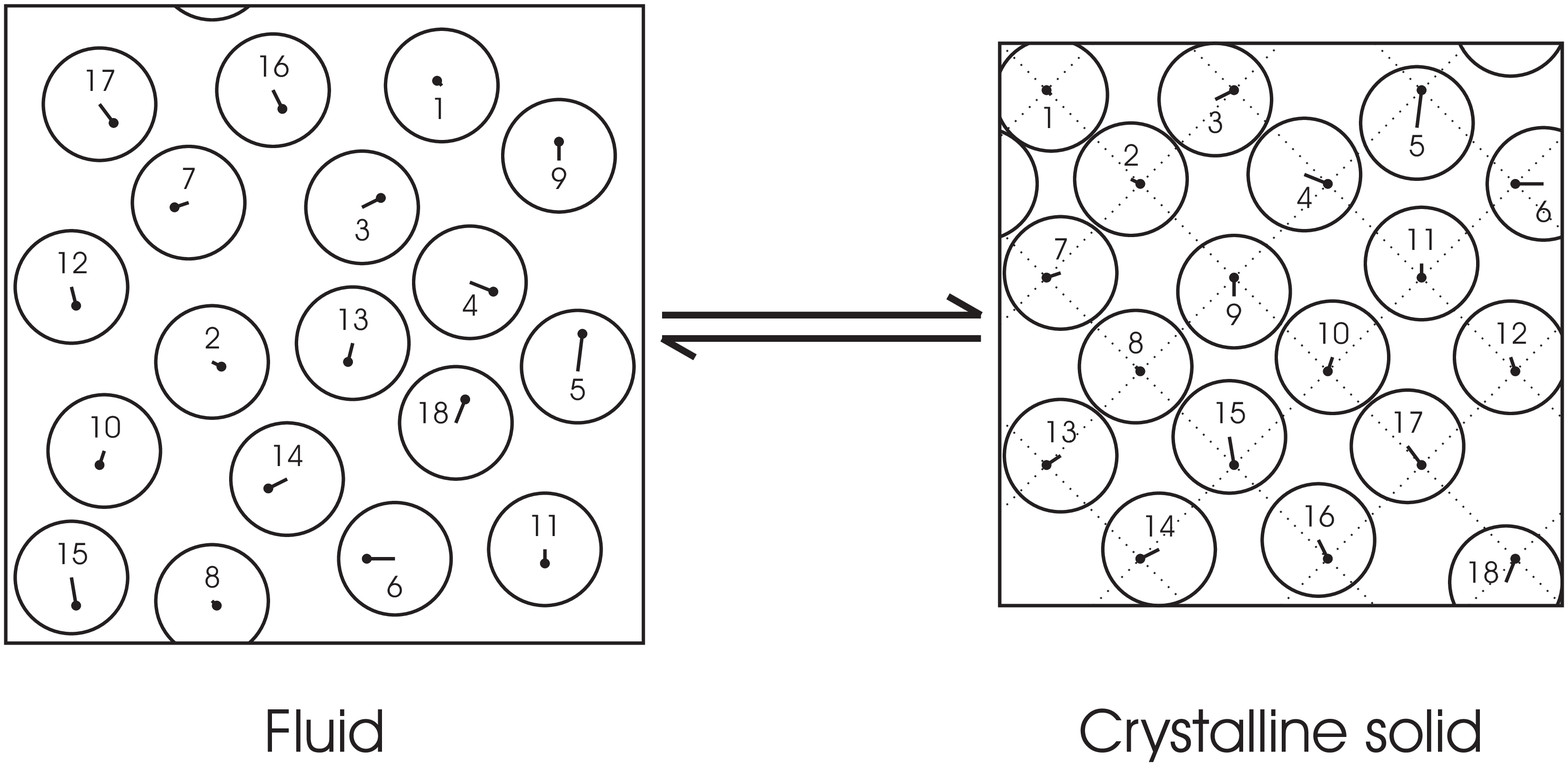}
\caption{Schematic illustration of the phase switch mechanism. The dots identify
the representative sites $\Rsetgamma$ in each phase; the displacement
vectors $\uset$ connect the centers of the distinguishable (numbered)
particles to these sites. The switch
operation shown swaps the representative sites of one phase for those
of the other phase, whilst maintaining $\uset$ constant. More
generally, any desired transformation of $\uset$ (eg. a scaling --see
text) can be incorporated in the switch \cite{BRUCE03}. The particular
configuration $\uset$ shown is a ``gateway'' state (see text) because the
magnitude of the effective energy change under the switch is small.}
\label{fig:psschem}
\end{figure}

The key to implementing the phase switch is the representation
(eq.~\ref{eq:origin}) of particle coordinates in term of displacements
with respect to a representative configuration $\Rsetgamma$. By
construction, the system may be transformed between the CS and F phase
representative configurations simply by switching the vectors
($\vec{R}_i^F \rightleftharpoons \vec{R}_i^{CS}\, \forall i$). Thus by
continuity, {\em any}~CS (F) configuration `sufficiently close' to the
representative one will also transform to a F (CS) state under this
operation \cite{glass}. This phase switch can itself be realized as an
MC step, so that the phase label $\gamma$ becomes a stochastic
variable. By `sufficiently close' here, we mean that the energy change
associated with the phase switch is such as to afford a reasonable
chance of acceptance. Fig.~\ref{fig:psschem} provides an
illustration of the mechanism. 

\begin{figure}[h]
\includegraphics[type=eps,ext=.eps,read=.eps,width=8.0cm,clip=true]{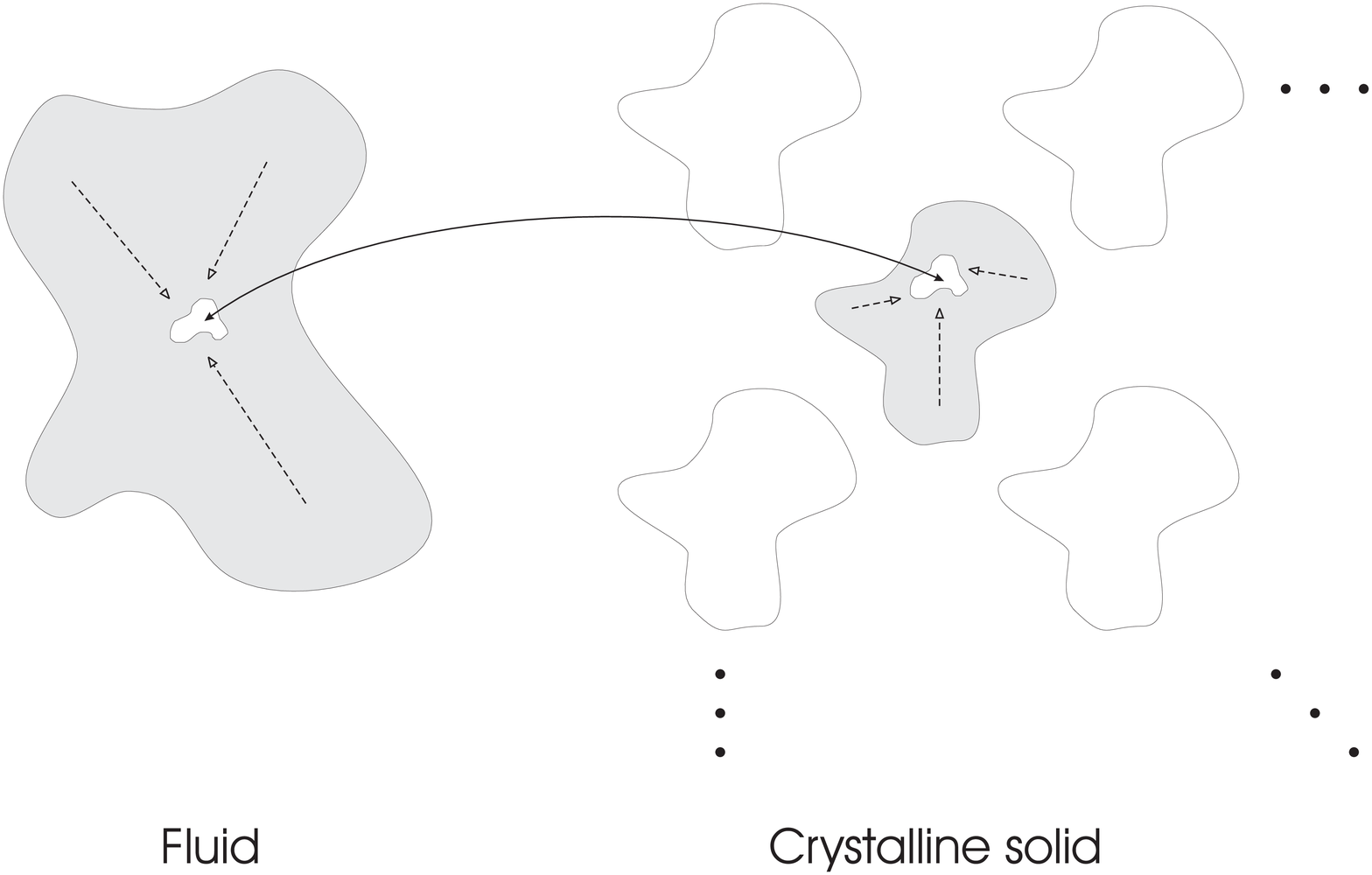}\\
\caption{Schematic illustration of the phase switch operation in terms of the
regions of configuration space associated with the F and CS phases. A
bias (dashed arrows) is constructed such as to enhance the probability of
the subsets of ``gateway'' states (the white islands) within the single-phase
regions, from which the switch operation (the large arrow) will be
accepted. Note that the switch accesses only one of the $(N-1)!$ CS
phase space replica fragments (see text).}
\label{fig:blobs}
\end{figure}

We term the set of configurations for which the MC phase switch will be
{\em accepted} the `gateway' states. In general, however, these gateway
states constitute only a small fraction of the total respective
configuration spaces of the phases, and consequently will rarely (if
ever) be visited on simulation timescales.  To ensure effective
two-phase sampling, the probabilities with which the gateway states are
visited must be {\em enhanced}. This can be achieved by appeal to
extended (biased) sampling methods.  Fig.~\ref{fig:blobs} shows a
schematic representation of the procedure. 

The bias necessary to promote sampling of the gateway states is
administered with respect to an `order parameter' --some suitably
chosen macrovariable of the system. The freedom in selecting this order
parameter is circumscribed by the requirement that the associated
microstates form a contiguous path through phase space linking the
large number of equilibrium (typical) microstates to the relatively
small number of gateway states. In the next subsection, we describe one
choice of order parameter which has proved itself successful in this regard.

\subsection{Order parameter and extended sampling strategy}
\label{sec:op}

One can devise a variety of order parameters that serve to form a
suitable path leading from the equilibrium states to the gateway states.
A definition that we have found to be effective is a variant of one
originally employed by one of us in ref.~\cite{WILDING00}, and which
closely resembles a recent proposal by Errington \cite{ERRINGTON04}.
Here the order parameter comes in two parts (or modes): `tether' and
`energy'. The tether mode serves to draw particles close to the
representative sites to which they are nominally associated. Then, once
all particles are within a prescribed distance of these sites, tether
mode switches off and an energy mode order parameter takes over to guide
particle to the gateway states for which the phase switch energy cost is
sufficiently small to be accepted.

To elaborate, let $M_{m,\gamma}$ denote the order parameter in mode $m$
and phase $\gamma$. Then for tether mode we write $m={\cal T}$ and
define an associated order parameter

\begin{equation}
\label{eq:tether}
M_{{\cal \tau},\gamma}(\uset)= \left( \frac{1}{N}\sum_{i=1}^N {\rm max}\{0,\tilde{u_i}-\tilde{u_c}\} \right)^{1/2}\:,
\end{equation}
where $\tilde{u_i}=|{\bf u}_i|/V^{1/3}$ is the distance of particle $i$
from its lattice site measured in units of the box length, and
$\tilde{u_c}$ is a prescribed dimensionless threshold radius. Only
particles whose displacement $\tilde{u_i}$ exceeds this threshold
contribute to $M_{{\cal T},\gamma}$.

The tether mode is active iff $\tilde{u}_i>\tilde{u_c}\:$ for at least one particle
$i$, i.e. when $M_{{\cal T},\gamma}>0$. Otherwise control hands over to the
`energy' mode $m={\cal E}$; its associated order parameter is defined by

\begin{equation}
M_{{\cal E},\gamma}(\uset)=\mathrm{sgn}(\Delta {\cal E}_{\gamma^\prime\gamma})\ln(1+|\Delta {\cal E}_{\gamma^\prime\gamma}|)\:,
\label{eq:Emode}
\end{equation}
where 

\begin{equation}
\label{eq:cost}
\Delta {\cal E}_{\gamma^\prime\gamma}=({\cal E}_\gamma-{\cal E}^{\rm ref}_\gamma)-  ({\cal E}_{\gamma'}-{\cal E}^{\rm ref}_{\gamma'}) 
\end{equation}
measures the change (under the phase switch operation) of the
Hamiltonian ${\cal E}_\gamma(\uset,V)=\Phi_\gamma(\uset)+pV$ with
respect to its value ${\cal E}^{\rm ref}$ in the 
representative microstate $\Rsetgamma$, with the latter scaled to the
instantaneous volume of phase $\gamma$ \cite{JACKSON,ERRINGTON04}. The
presence of the logarithm in eq.~\ref{eq:Emode} is designed to moderate
the scale of the contribution of the energy cost to $M_{{\cal
E},\gamma}$, which might otherwise become excessively large for
particles with a strongly repulsive core to their interaction potential.

Given these definitions, and recalling eqs.~\ref{eq:zsimpF} and
\ref{eq:zsimpCS}, the entire region of accessible configuration space
can be described by the ensemble

\begin{equation}
\zcaltilde {N} {p} {T} {\etaset} \equiv \sum _{\gamma} \int_{0}^{\infty} dV 
\prod_{i=1}^N\int_{\gamma} d \vec{u}_{i} e^{-{\cal H}_{\gamma}( \uset, V)}\:,
\label{eq:Ztot}
 \end{equation}
where ${\cal H}_\gamma$ is the {\em effective Hamiltonian} defined by 

\begin{eqnarray}
{\cal H}_{\gamma}( \uset ,  V) &=& \Phi( \uset , \Rsetgamma ) + p V +\eta_{m,\gamma}(M)\nonumber\\
& \:& -\delta_{\gamma,CS}\ln{(N-1)!}
\label{eq:effpot}
\end{eqnarray}
Here $\etaset$ represents a set of {\em multicanonical weights}
\cite{BERG,BRUCE03}, associated with the macrostates of the macrovariable $M=M_{m,\gamma}$.
The set can usefully be split into four `branches', one for each
combination of mode and phase.  The role of the weights is to modify
(with respect to the Boltzmann sampling) the probability with which the
various $M$ values are visited. More specifically, as discussed below,
we tailor their form in order to ensure roughly {\em uniform} sampling
over the relevant range of $M$.

Simulations in the ensemble described by eq.~\ref{eq:Ztot} allow
one to measure the multicanonical probability distribution
\begin{widetext}
\begin{equation}
\label{eq:sampdist}
P(M, V, {\cal E}, \gamma |N, p, T, \etaset )  \equiv \tilde{\cal Z}^{-1} \prod_{i=1}^N\int_{\gamma} d \vec{u}_{i} e^{-{\cal H}_{\gamma}( \uset , V)}\delta(M - M(\uset, V)) \delta( {\cal E}_\gamma - {\cal E}_\gamma(\uset, V)) \:,
 \end{equation}
\end{widetext}
the form of which is accumulated in practice via a simple list (see sec.~\ref{sec:data}).
The bias introduced by the multicanonical weights can be unfolded to give the 
equilibrium distribution (to within an unspecified normalization constant):

\begin{equation}
P(M, V,  {\cal E}, \gamma |N, p, T) \doteq  P(M, V,  {\cal E}, \gamma |N, p, T, \etaset) e^{\eta_{m,\gamma} (M)}
\end{equation}

The distribution of any single observable can be readily extracted from
$P(M, V, {\cal E}, \gamma)$ by integration. Similarly the ratio of
configurational weights is obtained as

\begin{eqnarray}
\label{eq:zratio}
\zratio &=&\frac {\zcal{F}{N}{p}{T}}   {\zcal{CS}{N}{p}{T}} \nonumber \\
        &=&\frac {\int dM\;dV\;d{\cal E}   \;P(M,V,{\cal E},F |N,p,T)}  {\int dM\;dV\;d{\cal E}\;  P(M,V,{\cal E},CS |N,p,T)} 
\end{eqnarray}
from which the Gibbs free-energy-density difference follows directly
via eq.~\ref{eq:gibbs}. 

\subsection{Implementation} 

\subsubsection{MC moves}
\label{sec:moves}

The Monte Carlo procedure we have adopted utilizes four types of
update, which we describe below. In implementing these updates, it is
computationally expedient to hold one particle {\em fixed} at its
representative site in each phase. This suppresses the global
translation mode in the CS phase and eliminates the need for
association swaps (see below) in this phase. Choosing to make the assignment
$\vec{r}_{i=N}\equiv  \vec{R}^{\gamma}_{i=N}$ (implying that $u_{i=N}=0$), we use
$\usetstar$ to represent sets $\uset$ of displacement coordinates which
satisfy this condition. Consequences for acceptance probabilities are
examined in appendix.~\ref{sec:append1}.

The MC procedure for one-phase exploration has three types of update:
`Particle translations', `Association swaps' and `Volume moves'. 

\begin{enumerate}

\item {\em `Particle translations'}. This is a standard MC procedure: a
site (identified by one of the vectors in $\RsetCS$ or $\RsetF$) is
chosen at random and a candidate state is chosen by incrementing the
position coordinate of the particle associated with that site by a
random vector whose components are drawn from a zero-mean uniform
distribution of prescribed width. This operation changes both $\rset$
and $\usetstar$

\item {\em `Association swaps'}. In this operation we choose two distinct sites
at random ($i$ and $j$ say) and identify the corresponding displacement
vectors $\vec{u}_i$ and $\vec{u}_j$.  The candidate state is defined by
the replacements

\begin{eqnarray}
\vec{u}_i &\rightarrow& \vec{u}_i^{\prime} \equiv\vec{u}_j + \vec{R}_j - \vec{R}_i \\
\vec{u}_j &\rightarrow& \vec{u}_j^{\prime} \equiv \vec{u}_i + \vec{R}_i - \vec{R}_j
\end{eqnarray}

This process can be regarded as a change of association: the
particle which was associated with site $j$ is now associated with
site $i$ (and {\it vice versa}).  It changes the {\em representation}
of the configuration (the coordinates $\uset$); but it leaves the
physical configuration invariant. 

We apply it only to the fluid phase where particles can wander far from
their representative sites and need to be reined back in order to reach
the gateway states. One {\em may} implement association
updates  in the CS phase too: the simulation can then be thought of as
exploring different CS-phase fragments;  the factor of $(N-1)!$ in
\protect Eq.~\ref{eq:zsimpCS}  is no longer needed and it is no longer
necessary to clamp one particle. 

\item {\em `Volume moves'}. The volume is also updated in the
conventional way, by a random walk of prescribed maximum step size,
with particle position coordinates $\{\bf{r}\}$ and representative sites $\{
\vec{R}\}^{\gamma}$ rescaled . Note that we maintain the ratio
$V_{\gamma^{\prime}}/V_{\gamma}$ constant throughout, so the notional
`conjugate' phase $\gamma^\prime$ also undergoes a dilation by the same
factor. 

\end{enumerate}

In each of the above cases the transition to the candidate state is accepted (see
appendix~\ref{sec:append1}) with the probability

\begin{eqnarray}
\lefteqn{p_a (\usetstar, V \rightarrow \usetstarprime, V^{\prime}\mid \gamma)=}\nonumber\\
& & {\rm min} \left\{1,\exp\left[-\Delta {\cal H}_\gamma+  N\ln(V^{\prime}/V) \right]\right\}
\end{eqnarray}

with 
\begin{equation}
\Delta{\cal H}_\gamma={\cal H}_{\gamma}( \usetstarprime , V^{\prime})-{\cal H}_{\gamma}( \usetstar, V)  \:.
\end{equation}
All three types of MC move may involve a change in the order parameter
$M$ and hence a change $\Delta \eta$ in the multicanonical weights
implicit in ${\cal H}$ (cf.~eq.~\ref{eq:effpot}). Note that particle
translations or association swaps that hand over control from one order
parameter mode to another, involve a weight change which is calculated
from the respective branches of the weight function. For example, for a
hand-over from tether mode to energy mode, in which the order parameter
changes from $M_{\cal T}$ to $M_{{\cal E}}$ the appropriate weight
change is calculated as 

\begin{equation}
\Delta \eta \equiv \eta_{{\cal E},\gamma}(M_{{\cal E},\gamma}) -\eta_{{\cal T}, \gamma}(M_{{\cal T},\gamma}) \:.
\end{equation}

\begin{enumerate}

\item[4] `{\em Inter-phase switch}'. The final type of MC update is the
phase switch, which entails replacing one set of the representative configuration
vectors, $\Rsetgamma$  say, by the other, $\Rsetgammaprime$.  If the
equilibrium densities of the two phases were close to one another  it
would also be possible to keep the volumes  constant in the switch. But
that is not the case in general. Accordingly --if the procedure is to
be efficient-- the switch needs to incorporate an appropriate {\em
volume scaling} of the system \cite{NOTE0}. It is natural to fix that scaling
so that a  `typical' volume $\hat{V}_\gamma$ of phase $\gamma$ is
matched to a `typical' volume $\hat{V}_{\gamma^\prime}$ of phase
$\gamma^\prime$. In practice we take $\hat{V}_{\gamma}$ to be the mean
volume of phase $\gamma$.  The switch is accepted (see
appendix~\ref{sec:append1})  with the probability 

\begin{eqnarray} 
\lefteqn{p_a (\gamma, \usetstar, V \rightarrow \gamma^{\prime}, \usetstarprime,  V^{\prime}) =}\nonumber \\
& &  \mbox{min} \left\{1,\exp\left[ -\Delta {\cal H}_{\gamma^\prime\gamma} + (N+1) \ln(\hat{V}_{\gamma^{\prime}}/\hat{V}_{\gamma})\right]\right\}
\end{eqnarray}

where 

\begin{equation}
\Delta {\cal H}_{\gamma^\prime\gamma}= {\cal H}_{\gamma^{\prime}}( \usetstarprime ,  V^{\prime}) -{\cal H}_{\gamma}(
\usetstar , V) 
\end{equation}

The phase switch occurs only from states in which the energy mode order
parameter is active; the associated change in multicanonical weights is

\begin{equation}
\Delta \eta \equiv \eta_{{\cal E},\gamma^\prime}(M^\prime_{{\cal E},\gamma^\prime})-\eta_{{\cal E},\gamma}(M_{{\cal E},\gamma}) \:,
\end{equation}
where $M^\prime_{{\cal E},\gamma^\prime}$=$-M_{{\cal E},\gamma}$ are
the new and old order parameter values respectively. This weight
discontinuity is at our disposal; we tune it so as to cancel  the
typical contribution to the acceptance probability associated with the
switch.

\end{enumerate}

When discussing our results (sec.~\ref{sec:results}) we shall
occasionally refer to a ``Monte Carlo sweep''. This we take to comprise
$N$ trial particle translations, $N$ trial association swaps, $1$ trial
volume move and $1$ trial phase switch.

\subsubsection{Data accumulation and histogram extrapolation}
\label{sec:data}

In the course of the simulations, observations are stored as lists
\cite{WILDING01} $V_j, M_j, {\cal E}_j, \gamma_j$ ($j= 1,2, \ldots$).
Estimates for the distribution of any observables were accumulated
by appropriate binning and weighting of the members of the list

\begin{equation}
P(X|N, p,T) \equalnorm \sum_j e^{\eta(M_j)} \delta(X-X_j)\:,
\end{equation}
with $X=V; M;$ etc.

In analyzing the data, extensive use was made of histogram
extrapolation (HE). This method greatly enhances computational
efficiency by facilitating the scanning of a region of pressure and
temperature around each simulation state point, without recourse to
further simulation.  To achieve this, HE {\em reweights} the data obtained at
one state point $(p,T)$ to yield estimates appropriate to another
not-too-distant state point $(p^\prime,T^\prime)$. The reliable range
of the extrapolation in $p$ and $T$ is dependent on the statistical
quality of the original data. For further information on the
HE technique, we refer the interested reader to ref.~\cite{HE}. Its
utility in the context of tracking a freezing line will be illustrated
in sec.~\ref{sec:results}.

\subsubsection{Weight function construction}

\label{sec:weights}

Prior to collection of phase switch data, it is necessary to determine
a suitable set of multicanonical weights $\etaset$ to enable the system
to pass readily between the equilibrium and gateway states. In practice
the requisite weights are defined with respect to the {\em macrostates}
formed by discretising (binning) the order parameter macrovariable
$M=M_{m,\gamma}$.  For a given mode $m$ and phase $\gamma$, the bin
width $\Delta M$ was chosen such that for all pairs of successive bins
$M^{(i)}$ and $M^{(i+1)}=M^{(i)}+\Delta M$, the corresponding weights
satisfy $\eta(M^{(i+1)})/\eta(M^{(i)})<2$. Doing so ensures that the
weight associated with any given macrostate is representative of the
full range of the underlying macrovariable, thereby guaranteeing a
reasonable MC acceptance rate.

Consideration of eqs.~\ref{eq:effpot} and eqs.~\ref{eq:sampdist} shows that
the form of the weights that confers equal probability on every
macrostate $M$ (and hence ensures that all are well sampled) is
$\eta(M)=-\ln P(M)$. For adequate sampling one therefore requires a
prior estimate of $P(M)$. A variety of techniques exist for providing
such an estimate, some of which are described in
refs.~\cite{BRUCE03,SMITH95}. The approach we have utilized in this
work is the transition matrix Monte Carlo method  (TMMC) 
\cite{SMITH95,FITZGERALD99,FITZGERALD00,ERRINGTON04}. 

TMMC extracts information on macrostate probabilities by focusing on
the statistics of {\em transitions} between these macrostates. The
method proceeds by defining a collection matrix, $\mathrm{C}$, in which
information regarding the macrostate transitions is accumulated. At
every MC step, the proposed transition is recorded in $\mathrm{C}$
(irrespective of whether or not it is actually accepted) according to:

\begin{eqnarray}
\mathrm{C}(M \!\to\! M') & \leftarrow \mathrm{C}(M\to M')& +\: a \nonumber\\ 
\mathrm{C}(M\!\to\! M) & \leftarrow \mathrm{C}(M\to M)& + \:(1 - a),
\label{eq:tmC}
\end{eqnarray}
where $a$ is the acceptance probability of the move under the Hamiltonian
$\mathcal{E}_\gamma(\{\mathbf{u}\}, V)$ (rather than the effective 
Hamiltonian $\mathcal{H}_\gamma$). 

The macrostate transition probabilities derive from the collection matrix via:

\begin{equation}
\mathrm{T}(M \!\to\! M') = \frac{\mathrm{C}(M \!\to\! M')}{\sum_{k} \mathrm{C}(M \!\to\! M_k)}\:,
\label{eq:tmP}
\end{equation}
where the denominator on the RHS sums over all possible values of the
macrostates. Knowledge of these transition probabilities permits, in
turn, an estimate of the probabilities $P(M)$. Although a number of
approaches exist (see eg. ref.~\cite{ERRINGTON04}), for extracting
$P(M)$ from T, a simple, yet unbiased and stable method, considers
exclusively those transitions which occur between adjacent macrostates
(i.e.~examines only the diagonal and first off-diagonal elements of
$\mathrm{T}$). Application of  the `balance' condition for
macrostate transitions \cite{SMITH95}:

\begin{equation}
\frac{P(M^\prime)}{P(M)} = \frac{\mathrm{T}(M \!\to\! M')}{\mathrm{T}(M' \!\to\! M)}\:,
\label{eq:tmDiff}
\end{equation}
then permits assignment of the weight function via

\begin{equation}
\eta(M^\prime)-\eta(M)=-\ln\left(\frac{\mathrm{T}(M \!\to\! M')}{\mathrm{T}(M' \!\to\! M)}\:,\right)\:.
\end{equation}

The weight function was updated in this manner at regular intervals
(typically every 1000 Monte Carlo sweeps). Each update allowed the
simulation to explore an ever wider range of order parameter values
\cite{ERRINGTON04}. We note, however, that weight function updates
should not be performed too frequently during the weight evolution
process because, strictly speaking, they violates detailed balance.
Empirically, however, at the stated update interval, the solution was
found to be highly accurate and the procedure efficient.

For the purposes of constructing the weight function, we found it
beneficial to decompose the full range of $M$ into a number of
overlapping `windows'. Within each window a fragment of the overall
weight function was determined and joined for continuity to those of
neighboring windows. The weight differences across the three mode
transitions (tether to energy mode for each phase, as well as the phase
switch itself) were found using a simple root-finding algorithm in a
simulation restricted to the appropriate extremes of the order
parameter. Both energy modes were calculated as single windows, while
the solid and fluid tether modes were split into 3 and 10 windows
respectively (for the largest system size studied $N=500$). 

\section{Application to the Lennard-Jones system}

\label{sec:results}

\subsection{Model and strategy}
\label{sec:model} 

We have employed the Phase Switch Monte Carlo method to study the
freezing properties of particles whose interactions are described by 
the Lennard-Jones potential

\begin{equation}
\Phi(r)=4\epsilon\left[\left(\frac{\sigma}{r}\right)^{12}-\left(\frac{\sigma}{r}\right)^{6}\right]\:.
\label{eq:lj}
\end{equation}
The system is assumed to be contained within a cubic box of volume $V$,
which fluctuates under the control of an applied external pressure $p$
and is in thermal equilibrium at temperature $T$. In common with
previous studies of freezing in the LJ system
\cite{AGRAWAL95,ERRINGTON04}, particle interactions were truncated at
one-half the box length, and running tail corrections applied to the
energy in the standard manner \cite{FRENKELSMIT}. As described in
sec.~\ref{sec:moves}, the MC moves employed were particle translations,
volume moves, association swaps and phase switches. Of these, we note
that  volume moves and phase switches are computationally inexpensive
to implement because the associated energy change is obtainable
directly from the scaling with (respect to coordinate dilations) of the
repulsive and attractive parts of the configurational energy
\cite{ALLEN}. Association swaps are similarly cheap because they leave
the configurational energy unaltered. Since the position of one
particle was clamped at its representative site in each phase, CS phase
association swaps were not required (see sec.~\ref{sec:moves} and
appendix~\ref{sec:append1}). 

The phase switch MC method requires for its operation the specification
of a representative configuration $\Rsetgamma$ for both the CS and F
phases. For $\RsetCS$ we adopt a perfect lattice of the appropriate
symmetry (here fcc) and scale (corresponding to a rough estimate of the
CS phase coexistence density). For $\RsetF$, we randomly select a
configuration in the course of a simulation of the fluid phase
\cite{NOTE1}. 

With regard to the order parameter (sec sec~\ref{sec:op}), the
hand-over from tether to energy mode is controlled by the parameter
$\tilde{u_c}$ in eq.~\ref{eq:tether}. This should be chosen such
that --at the hand-over-- particles are sufficiently close to their
representative sites that the energy cost $\Delta {\cal
E}_{\gamma^\prime\gamma}$ of the switch (eq.~\ref{eq:cost}) is not
saturated at a large value. Otherwise $M_{{\cal E},\gamma}$ will be
unable to guide the particles to positions favorable for the phase
switch. We find that satisfactory results are obtained by choosing
$\tilde{u_c}$ consistent with $M_{{\cal E},\gamma}\approx 5$ at the
hand-over.

The formalism described in Secs.~\ref{sec:statmech} and
~\ref{sec:method} assumes that the configurational constraint confines
the CS phase to the phase space fragment in which it is initiated. While
this appears to be strictly observed for F-CS coexistence in hard
spheres, as studied in ref.~\cite{WILDING00}, for the softer
interactions of the LJ system, interchange of neighboring particles
between lattice sites was observed to occur, albeit very rarely. The
simplest way of dealing with this effect is to suppress it; to which end
we introduced an upper bound on the particle displacements in the CS
phase. We set $|u_i|^{CS}\le 0.65\sigma$, a choice which (we have
verified) has a negligible effect on the equilibrium properties of the
CS phase for the system sizes we have studied.

\begin{figure}[h]
\includegraphics[width=8.0cm,clip=true]{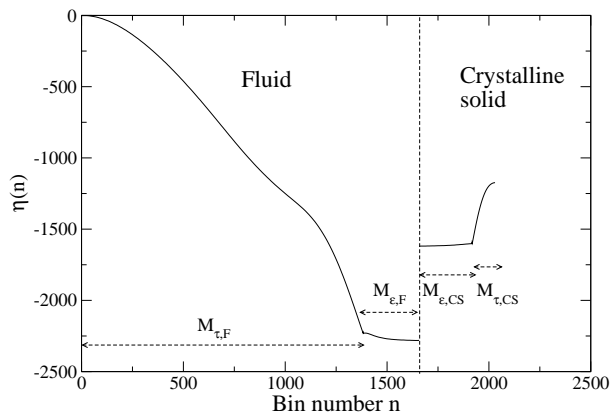}
\caption{An example weight function used for a near-coexistence production run. The data shown are for a system of
$N=256$ LJ particles at $\beta=0.6$, $P=13.7$. The four branches of the
weight function have been laid out with respect to the underlying
discretisation into bins (see sec.~\ref{sec:weights}).}
\label{fig:exampleweight}
\end{figure}

The strategy for obtaining the F-CS coexistence line is as follows.
Initially one requires a rough estimate for some coexistence state
point. This may derive either from any available literature estimates
or, alternatively, coexistence can be bracketed by scanning (backwards 
and forwards) a range of pressures at some nominated temperature until the
system spontaneously transforms between phases. Next it is necessary to
customize a suitable set of multicanonical weights which allow the
system to sample both phases. This can be done by employing, for
example, the transition matrix method described in
sec.~\ref{sec:weights}. An example weight function is shown in
fig.~\ref{fig:exampleweight} for a system of $N=256$ particles close to
coexistence at $\beta=\epsilon/kT=0.6$. 

\begin{figure}[h]
\includegraphics[width=8.0cm,clip=true]{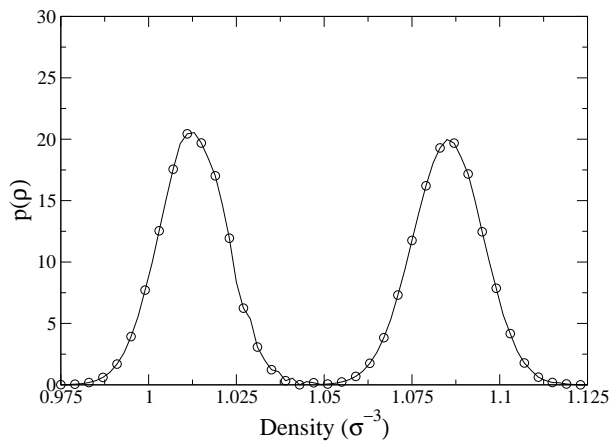}
\caption{The coexistence form of the density distribution $P(\rho)$ for
a system of $N=256$ LJ particles at $\beta=0.6$, $p=13.722$. A
selection of data points are shown. Lines are guides to the eye.}
\label{fig:density_hist}
\end{figure}

Having determined a suitable weight function, a production run is then
performed to obtain data of high statistical quality. From this data
one can readily extract (see sec.~\ref{sec:data}) the ratio
$\zratio$ (eq.~\ref{eq:zratio}) and hence the
free energy difference $\Delta g$ (eq.~\ref{eq:gibbs}) between the
phases. Histogram extrapolation (HE) (sec.~\ref{sec:data}) then permits
the location of the coexistence pressure at which this difference
vanishes (i.e. when $\zratio=1$).  The corresponding coexistence number
densities can then be simply read off from the peak positions in the
coexistence form of $P(\rho)$, where $\rho=N/V$.
Fig.~\ref{fig:density_hist} shows a typical coexistence form of
$P(\rho)$, obtained for $N=256$ at $\beta=0.6$.

In addition to simplifying the determination of the coexistence
pressure at a given temperature, HE also permits simultaneous
extrapolation in the temperature. This facilitates the scanning of a
segment of the coexistence line in the $p-T$ plane. Although the range
of near coexistence $p-T$ values for which a single simulation 
provides reliable information is necessarily limited, additional
simulations can be performed near the vestiges of this range. The
data from simulations performed at a number of different state
points can then be synthesised self-consistently using multiple histogram
reweighting \cite{HE}. As well as guiding the choice of near
coexistence values of $p$ and $T$ for subsequent simulations,
HE also provides an estimate of the appropriate order
parameter distribution $P(M_{m,\gamma})$, which (recall
sec.~\ref{sec:weights}) serves as a suitable weight function. Thus by a
repeated combination of simulation followed by HE, the coexistence line
can be straightforwardly tracked, without the need to ever re-determine
a weight function {\em ab initio} \cite{WILDING01}. 

\subsection{Results}

Estimates of the freezing boundary have been obtained using the PSMC method
within the constant-$NpT$ ensemble for systems of sizes $N=108, 256$ and
$500$ particles. Freezing boundary data has also been obtained for
$N=32$, but here the system is sufficiently small that transitions back
and forth between F and CS phases occur {\em spontaneously}, over a
range of pressures, and a density distribution (sampling both phases)
can be determined --and a coexistence pressure inferred-- without the
need for PSMC.

Spontaneous transitions were also found to occur (albeit much more
rarely) in the $N=108$ system. Here we observed spontaneous freezing as
we tracked the coexistence curve using the methods described in
sec.~\ref{sec:model}.  We emphasize that this effect was also seen for
this system size in a `bare' constant-$NpT$ simulation at coexistence
and does not therefore appear to be caused (or rendered more frequent)
by the multicanonical weighting of PSMC. The larger system sizes
($N=256$ and $N=500$) did not exhibit spontaneous transitions on the
timescale of our simulations,  presumably because of the increased free
energy cost of forming an interface. 

The spontaneous freezing that occurs in the $N=108$ system undermines
our strategy whereby the phase label $\gamma$ keeps track of the
current phase. Nevertheless, we were able to suppress the problem by
tracking the coexistence line along a path parallel to its true
trajectory, but displaced slightly towards the fluid side. This was
achieved by simulating at a pressure some $5\%$ less than the
coexistence value predicted by the histogram extrapolation from the
previous near-coexistence state point. The estimate of the coexistence
pressure was easily corrected for the imposed shift by means of
histogram extrapolation.

\begin{figure}[h]
\includegraphics[width=8.5cm,clip=true]{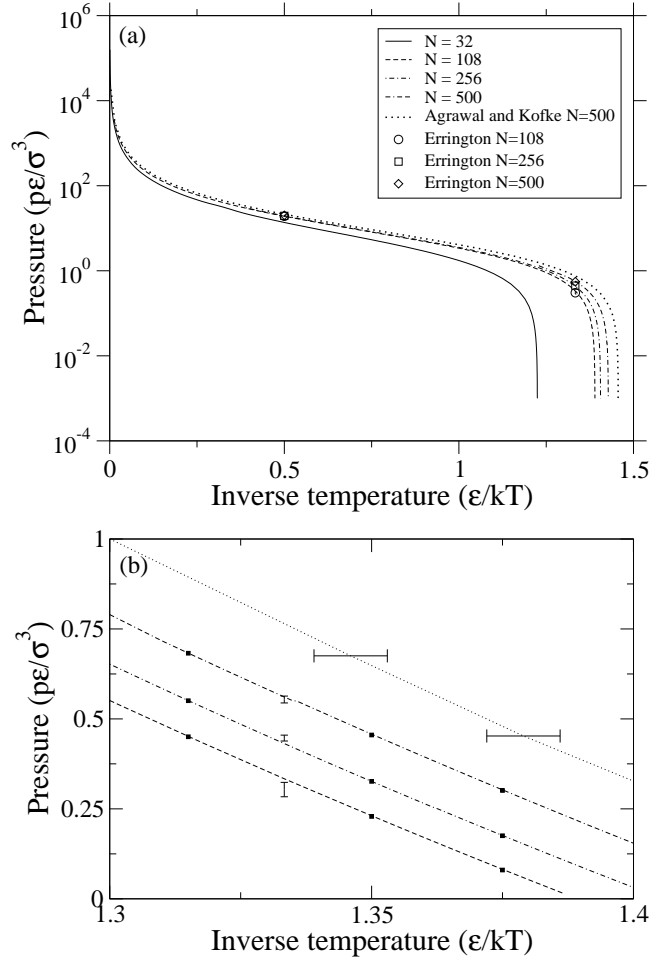}

\caption{{\bf (a)} The estimated F-CS phase diagram in the
pressure-inverse temperature plane for the four systems sizes studied in
this work. The data shown derive from $20$ separate simulation state
points for the $N=32$ system size, $37$ state points for $N=108$, $17$
points for $N=256$ and $4$ points for the $N=500$ system size. Also
included for comparison are the PSMC estimates of Errington
\cite{ERRINGTON04} for $\beta=4/3, 1/2$, together with the GDI results
of Agrawal and Kofke \cite{AGRAWAL95} for $N=500$. {\bf (b)} A closeup
of the region around $\beta=4/3$. The vertical error bars corresponding
to Errington's data points \cite{ERRINGTON04} and the horizontal error
bars to the GDI study. Symbols are the estimates arising from the
present study (given explicitly in
tables~\ref{tab:LJpd108}-\ref{tab:LJpd500}), with uncertainties smaller
than the symbol size in each case. Lines are interpolations between the
data points, based on histogram extrapolation.}

\label{fig:LJpd}
\end{figure}

\begin{figure}[h]
\includegraphics[width=8.5cm,clip=true]{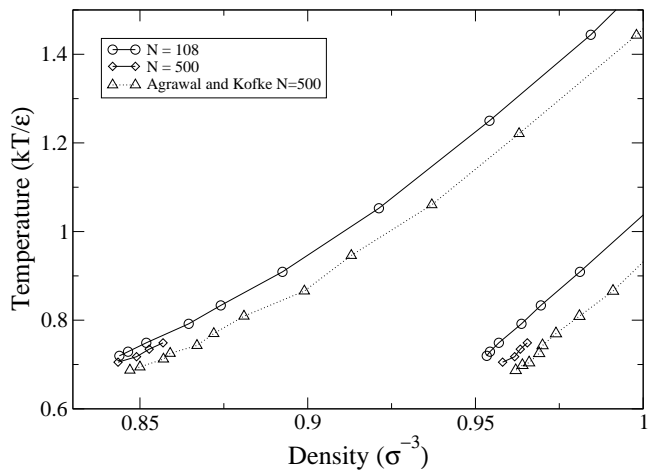} 
\caption{A portion of the phase diagram in the $\rho-T$ plane in the
region of the triple point. Shown are the estimated solid and fluid coexistence
densities for $N=500$ and $N=108$. Also included for comparison are the GDI
estimates of Agrawal and Kofke\cite{AGRAWAL95} for $N=500$.
Uncertainties are smaller than the symbol sizes; lines are
guides to the eye} 
\label{fig:pd-pT} 
\end{figure}

The measured phase boundary in the $p-T$ plane, and a portion of the
phase diagram in the $\rho-T$ plane, are shown in figs.~\ref{fig:LJpd}
and ~\ref{fig:pd-pT} respectively for the various system sizes studied.
Full data are also tabulated in
tables~\ref{tab:LJpd108}-\ref{tab:LJpd500} together with their
uncertainties. With regard to the latter, we note that
analysis of statistical errors within the PSMC framework is extremely
transparent: at a given temperature, the uncertainty  in the pressure is
simply given by 

\begin{equation}
\sigma[p]=\frac{\sigma[{{\mathcal{R}}_{\mbox{\sc {f,cs}}}}]}{|\Delta \bar{V}|}\:, 
\label{eq:error}
\end{equation}
where $\Delta \bar{V}=\bar {V}_F-\bar {V}_{CS}$, and
$\sigma[{{\mathcal{R}}_{\mbox{\sc {f,cs}}}}]$ is the associated uncertainty in
$\zratio$, this being controlled (at heart) by the
statistics of the inter-phase switch and measurable from a simple block
error analysis. We note that owing to the extensive factor $\Delta
\bar{V}$ in the denominator of eq.~\ref{eq:error}, the {\em minimum} uncertainty
in $\zratio$ required to attain a prescribed error in
the pressure {\em grows} like $N$  \cite{ERRORSNOTE}. 

As with any method, the error analysis becomes more involved if one
chooses to combine a number of separate data sets via a multiple
histogram analysis \cite{HE}. In these circumstances we have found it
useful to perform a bootstrap error analysis with $100$ re-samples,
taking an estimate of the block size as the correlation length. The
error estimate was found to depend only weakly on the block size. All
error bars in tables~\ref{tab:LJpd108}-\ref{tab:LJpd500} are calculated
in this way and correspond to a $67 \%$ confidence interval. We stress
that it is not strictly necessary to perform this more sophisticated
error analysis,  adequate (albeit slightly overly conservative) error
bars can be assigned simply on the basis of block averages.

\begin{table}
\begin{tabular}{r@{}lr@{}lr@{}lr@{}lr@{}lr@{}l}
\hline
\multicolumn{12}{c}{\rule{0pt}{9pt}$N = 108$} \\
\multicolumn{2}{c}{$\epsilon/kT$} & \multicolumn{2}{c}{$p\sigma^3/\epsilon$} & \multicolumn{2}{c}{$\rho_{CS}\,\sigma^3$} & \multicolumn{2}{c}{$\rho_F\,\sigma^3$} & \multicolumn{2}{c}{$\mathcal{E}_{CS}/N\epsilon$} & \multicolumn{2}{c}{$\mathcal{E}_F/N\epsilon$} \\[2pt]
\hline
0&.01250     & 3489&(8)   & 2&.5682(18) & 2&.5531(10) & 1563&.4(6)  & 1572&.15(31) \\
0&.01550     & 2635&(9)   & 2&.4141(11) & 2&.3277(13) & 1204&.7(3)  & 1251&.84(54) \\
0&.01900     & 2039&(3)   & 2&.3045(4)  & 2&.2197(2)  & 1008&.5(1)  & 1050&.32(8)  \\
0&.02350     & 1549&(2)   & 2&.1905(2)  & 2&.1091(1)  & 813&.48(7)  & 847&.951(32) \\
0&.02800     & 1232&(1)   & 2&.1014(3)  & 2&.0212(5)  & 632&.69(7)  & 660&.221(73) \\
0&.03500     & 922&(1)    & 1&.9935(2)  & 1&.9182(3)  & 503&.94(5)  & 526&.283(37) \\
0&.04500     & 663&.1(9)  & 1&.8803(2)  & 1&.8078(3)  & 389&.12(4)  & 406&.995(38) \\
0&.05500     & 507&.7(8)  & 1&.7935(1)  & 1&.7235(2)  & 312&.12(2)  & 326&.930(22) \\
0&.06500     & 406&.5(6)  & 1&.7260(2)  & 1&.6576(1)  & 257&.63(2)  & 270&.276(10) \\
0&.07500     & 334&.6(9)  & 1&.6702(7)  & 1&.6033(3)  & 219&.03(6)  & 230&.064(21) \\
0&.09000     & 262&.0(7)  & 1&.6026(7)  & 1&.5383(5)  & 169&.65(3)  & 178&.562(29) \\
0&.11000     & 197&.7(4)  & 1&.5299(3)  & 1&.4659(2)  & 135&.32(2)  & 142&.881(16) \\
0&.13000     & 157&.8(3)  & 1&.4773(3)  & 1&.4142(2)  & 110&.82(2)  & 117&.339(8)  \\
0&.16000     & 117&.2(2)  & 1&.4111(4)  & 1&.3477(6)  & 82&.416(15) & 87&.742(24)  \\
0&.17500     & 103&.0(3)  & 1&.3827(3)  & 1&.3213(5)  & 72&.448(1)  & 77&.294(24)  \\
                                                                                     
0&.21000     & 78&.95(13) & 1&.3304(3)  & 1&.2692(1)  & 56&.230(9)  & 60&.335(4)   \\
0&.24000     & 64&.68(8)  & 1&.2940(1)  & 1&.2327(1)  & 47&.043(4)  & 50&.738(5)   \\
0&.30000     & 45&.66(6)  & 1&.2349(1)  & 1&.1734(4)  & 30&.560(4)  & 33&.483(7)   \\
0&.37000     & 32&.43(5)  & 1&.1835(1)  & 1&.1214(1)  & 22&.669(2)  & 25&.210(3)   \\
                                                                                     
0&.43000     & 25&.12(4)  & 1&.1505(11) & 1&.0863(4)  & 16&.44(50)  & 18&.718(10)  \\
0&.50000     & 19&.13(3)  & 1&.1180(11) & 1&.0525(3)  & 11&.17(34)  & 13&.163(6)   \\
0&.62000     & 12&.56(4)  & 1&.0762(12) & 1&.0070(3)  & 5&.21(16)   & 6&.908(4)    \\
0&.69257     & 9&.840(17) & 1&.0557(11) & 0&.9844(2)  & 2&.72(8)    & 4&.271(3)    \\
0&.80000     & 6&.859(8)  & 1&.0291(11) & 0&.9542(2)  & -0&.23(1)   & 1&.157(3)    \\
0&.95000     & 4&.105(10) & 1&.0023(11) & 0&.9212(1)  & -2&.98(91)  & -1&.735(2)   \\
1&.10000     & 2&.236(7)  & 0&.9812(11) & 0&.8925(2)  & -4&.85(15)  & -3&.701(2)   \\
1&.20000     & 1&.293(4)  & 0&.9695(11) & 0&.8741(1)  & -5&.86(18)  & -4&.752(1)   \\
1&.26282(42) & 0&.8173    & 0&.9637(11) & 0&.8645(1)  & -6&.30(19)  & -5&.207(1)   \\
1&.33510(54) & 0&.3150    & 0&.9570(11) & 0&.8519(2)  & -6&.83(21)  & -5&.755(1)   \\
1&.37193(39) & 0&.1003    & 0&.9543(11) & 0&.8464(3)  & -7&.07(21)  & -5&.994(2)   \\
1&.38077(56) & 0&.0466    & 0&.9537(11) & 0&.8447(2)  & -7&.13(21)  & -6&.051(2)   \\
1&.38564(46) & 0&.0250    & 0&.9537(11) & 0&.8447(2)  & -7&.15(22)  & -6&.080(1)   \\
1&.38674(47) & 0&.0150    & 0&.9533(11) & 0&.8438(2)  & -7&.16(22)  & -6&.087(2)   \\
1&.38963(91) & 0&.0010    & 0&.9534(11) & 0&.8439(1)  & -7&.18(22)  & -6&.105(1)  \\
\hline
\end{tabular}
\caption{Solid-fluid coexistence curve data for the $N=108$ system.
Tabulated in columns $1-6$, respectively, are the inverse freezing
temperature, pressure, solid number density, fluid number density,
energy per particle (solid) and energy per particle (fluid). Numbers
in parentheses indicate the the $67\%$ confidence limit for the
rightmost digit(s). Note that due to the steepness of the coexistence
curve at low pressures (cf.~fig.~\ref{fig:LJpd}), it is expedient in
this regime to determine the coexistence temperature at prescribed
pressure.}

\label{tab:LJpd108}
\end{table}

\begin{table}[h]
\begin{tabular}{r@{}lr@{}lr@{}lr@{}lr@{}lr@{}l}
\hline
\multicolumn{12}{c}{\rule{0pt}{9pt}$N = 256$} \\
\multicolumn{2}{c}{$\epsilon/kT$} & \multicolumn{2}{c}{$p\sigma^3/\epsilon$} & \multicolumn{2}{c}{$\rho_{CS}\,\sigma^3$} & \multicolumn{2}{c}{$\rho_F\,\sigma^3$} & \multicolumn{2}{c}{$\mathcal{E}_{CS}/N\epsilon$} & \multicolumn{2}{c}{$\mathcal{E}_F/N\epsilon$}\\[2pt]
\hline
0&.39000     & 30&.400(47)  & 1&.1771(3) & 1&.10965(3)  & 20&.200(22)  & 22&.8182(6)  \\
0&.43000     & 25&.390(180) & 1&.1525(5) & 1&.08401(12) & 15&.870(22)  & 18&.2534(30) \\
0&.47000     & 21&.942(37)  & 1&.1355(2) & 1&.06611(1)  & 12&.430(13)  & 14&.6269(1)  \\
0&.54000     & 16&.928(34)  & 1&.1065(2) & 1&.0355(2)   & 8&.1409(89)  & 10&.0970(21) \\
0&.60000     & 13&.667(49)  & 1&.0854(3) & 1&.0130(2)   & 5&.4899(69)  & 7&.2971(19)  \\
0&.68000     & 10&.470(23)  & 1&.0616(2) & 0&.9861(5)   & 3&.1561(39)  & 4&.8445(52)  \\
                                                                                        
0&.68000     & 10&.477(18)  & 1&.0619(4) & 0&.9864(1)   & 3&.1494(43)  & 4&.8415(15)  \\
0&.80000     & 7&.068(12)   & 1&.0332(3) & 0&.9532(1)   & -0&.0465(6)  & 1&.4633(7)   \\
0&.91000     & 4&.885(13)   & 1&.0126(3) & 0&.9278(1)   & -2&.1525(28) & -0&.7639(17) \\
1&.00000     & 3&.546(9)    & 0&.9990(3) & 0&.9098(1)   & -3&.5496(44) & -2&.2385(8)  \\
1&.10000     & 2&.340(9)    & 0&.9851(3) & 0&.8901(1)   & -4&.7187(56) & -3&.4660(11) \\
1&.20000     & 1&.418(6)    & 0&.9745(3) & 0&.8734(1)   & -5&.7189(62) & -4&.5181(11) \\
1&.29969(94) & 0&.6456      & 0&.9651(3) & 0&.8560(3)   & -6&.5252(70) & -5&.3518(26) \\
1&.33481(74) & 0&.4263      & 0&.9628(3) & 0&.8508(1)   & -6&.7631(74) & -5&.5961(9)  \\
1&.36274(71) & 0&.2500      & 0&.9604(3) & 0&.8463(2)   & -6&.9512(74) & -5&.7908(18) \\
1&.38853(64) & 0&.1000      & 0&.9587(3) & 0&.8422(2)   & -7&.1148(76) & -5&.9582(11) \\
1&.40025(60) & 0&.0300      & 0&.9578(3) & 0&.8405(1)   & -7&.1896(77) & -6&.0372(8)  \\
1&.40527(68) & 0&.0010      & 0&.9574(3) & 0&.8395(1)   & -7&.2208(77) & -6&.0681(6)  \\
\hline
\end{tabular}
\caption{Solid-fluid coexistence curve data for $N=256$. See caption of tab.~\protect\ref{tab:LJpd108} for details.}
\label{tab:LJpd256}
\end{table}

\begin{table}[h]
\begin{tabular}{r@{}lr@{}lr@{}lr@{}lr@{}lr@{}l}
\hline
\multicolumn{12}{c}{\rule{0pt}{9pt}$N = 500$} \\
\multicolumn{2}{c}{$\epsilon/kT$} & \multicolumn{2}{c}{$p\sigma^3/\epsilon$} & \multicolumn{2}{c}{$\rho_{CS}\,\sigma^3$} & \multicolumn{2}{c}{$\rho_F\,\sigma^3$} & \multicolumn{2}{c}{$\mathcal{E}_{CS}/N\epsilon$} & \multicolumn{2}{c}{$\mathcal{E}_F/N\epsilon$} \\[2pt]
\hline
1&.336(2) & 0&.5500 & 0&.9655(1) & 0&.8566(2) & -6&.637(1) & -5&.486(2) \\
1&.3613(3) & 0&.3836 & 0&.96337(4) & 0&.85274(3) & -6&.815(1) & -5&.6705(2) \\
1&.393(1) & 0&.2091 & 0&.9617(1) & 0&.8490(1) & -7&.007(1) & -5&.871(1) \\
1&.418(2) & 0&.0200 & 0&.9581(1) & 0&.843(1) & -7&.198(1) & -6&.08(1) \\
\hline
\end{tabular}
\caption{Solid-fluid coexistence curve data for $N=500$. See caption
of tab.~\protect\ref{tab:LJpd108} for details.}

\label{tab:LJpd500}
\end{table}

Ideally one should like to extrapolate coexistence parameters obtained
for a range of finite system sizes to the thermodynamic limit using a
finite-size scaling ansatz. In the hard sphere case, a clear $N^{-1}$
dependence of the coexistence pressure has been observed using the PSMC
method \cite{WILDING00,ERRINGTON04}, and there seems no reason to
expect a different scaling in the case of the LJ potential.
However, in the present study (as well as in that of
ref.~\cite{ERRINGTON04}), a good fit to this scaling form could not be
obtained. As noted in ref.~\cite{ERRINGTON04}, the reason for this is
presumably traceable to the truncation of the interparticle potential
cutoff, which was set to be one-half the box length in all cases. This
choice, while facilitating direct comparison with existing literature
data (principally ref.~\cite{AGRAWAL95}), necessarily introduces a
degree of coupling between the potential truncation and the system
size, notwithstanding the use of a mean-field based tail correction.
For the rather limited range of system sizes studied in this work, the
use of such a tail correction is unlikely to provide a good
approximation to the full potential, particularly for the smaller
system sizes. Indeed, it is well known that many aspects of the phase
behaviour of the LJ system are acutely sensitive to the details of the
potential cutoff \cite{JACKSON,SMIT92,ERRINGTON04}.

\section{Discussion and conclusions}
\label{sec:concs}

In summary, the Phase Switch MC method allows one to locate --directly
and transparently-- fluid-solid coexistence parameters and their
associated uncertainties within the appropriate (constant-pressure)
ensemble. To achieve this the method connects the phases via a direct
inter-phase sampling path, thereby facilitating estimates of free
energy differences from a single simulation. The course of this path
encompasses configurations which are exclusively pure-phase in character.
Accordingly, one can be confident that it robust, i.e. free of both
singularities and crystalline defects.

We have applied the PSMC method to calculate, for a number of system
sizes, significant portions of the $p-T$ freezing curve of the LJ fluid
to high accuracy (see the error bars quoted in
tables~\ref{tab:LJpd108}-\ref{tab:LJpd500}). In so doing we have
illustrated the strategic advantages of histogram extrapolation (HE) as an
efficient means of tracking the coexistence line. Furthermore we have
seen (cf.~refs.~\cite{WILDING95,WILDING01}) that HE provides reliable
estimates of the requisite multicanonical weight function, thereby
obviating the need to redetermine the weights from scratch at each
successive near-coexistence state point. While it is probably fair to
say that the PSMC remains a somewhat computationally intensive strategy
(at least in the context of the F-CS problem), it is by no means
prohibitively so on the scale of its competitors. Given the strengths
of the method we have identified, we feel it should represent an
attractive option, especially when high precision estimates of freezing
boundaries are required.

We have compared our results with those of the previous PSMC study of
Errington \cite{ERRINGTON04} and the GDI study of Agrawal and Kofke
\cite{AGRAWAL95} for $N=500$. Not surprisingly, our results agree (to
within error) with those of Errington. However, (and as also noted by
Errington), the results for $N=500$ do not overlap with the error bars
quoted in the GDI study of the same system size (and potential
truncation) by Agrawal and Kofke \cite{AGRAWAL95}. This is possibly due
to the inherent limitations of the GDI scheme as already noted in
sec.~\ref{sec:intro} --specifically the lack of any means to reconnect
the phases beyond the initial `starting' coexistence state point from
where the integration is launched. It may additionally reflect a poor
estimate of the starting point itself, which was determined using TI by
transforming the LJ system to a hard sphere reference system via a route
which takes in an intermediate system of soft spheres. In particular,
the coexistence pressure taken for the hard sphere reference system is
now believed to be an overestimate
\cite{WILDING00,ERRINGTON04,SWEATMAN05}.

As regards the prospects for further applications, it would certainly
be worthwhile in the specific case of the LJ system to attempt to
decouple the system size scaling of the cutoff from intrinsic
finite-size effects by repeating the present study for a constant
interparticle potential cutoff. This should permit a reliable
extrapolation of coexistence parameters to the thermodynamic limit,
albeit for the particular choice of cutoff. More generally, it would be
interesting to attempt to apply the PSMC method to molecular systems.
Here, however, the order parameter would probably need to be
augmented in order to not only draw molecules to their representative
sites, but also to align them appropriately.

\section*{Acknowledgement}

The authors thank Jeff Errington for providing numerical data
from ref.~\protect\cite{ERRINGTON04}. 

\appendix

\renewcommand{\theequation}{\Alph{section}\arabic{equation}}

\section{Acceptance criteria}
\label{sec:append1}

Here we derive the Monte Carlo acceptance probabilities used in the
PSMC method.

As described in the sec.~\ref{sec:method}, we choose to fix one
particle at its representative site. The effect of this is to
reduce the configurational weight of each phase by a factor of the
system volume $V$. Thus for the fluid phase, eq.~\ref{eq:zsimpF}
becomes

\begin{equation}
\zsimp{F}{N}{V}{T} = V\prod_{i=1}^{N-1}\int_{V,F} d \vec{u}_i
e^{-\Phi( \usetstar , \RsetF )} \:,
\end{equation}
while for the CS phase (using eq.~\ref{eq:zsimpCS}), we have

\begin{equation}
\zsimp{CS}{N}{V}{T} = (N-1)! V \prod_{i=1}^{N-1}\int_{V,CS} d \vec{u}_i
e^{-\Phi( \usetstar , \RsetCS) }\:.
\end{equation}

In analogy to eq.~\ref{eq:Ztot} the phase space relevant to the problem
is described by the multicanonical ensemble 

\begin{equation}
\label{zcaltilde}
\zcaltilde {N} p T {\etaset} \equiv
\sum _{\gamma} \int_{0}^{\infty} V dV  \prod_{i=1}^{N-1}\int_{\gamma} d \vec{u}_{i} 
e^{-{\cal H}_{\gamma}( \usetstar , V)}
 \end{equation}
where
\begin{eqnarray}
{\cal H}_{\gamma}( \usetstar ,  V) &=&\Phi( \usetstar , \Rsetgamma ) +p V+\nonumber\\
&\: &   \eta_{m,\gamma} (M) - \delta_{\gamma,CS}\ln{(N-1)!}
\end{eqnarray}

The configurational probability follows as

\begin{equation}
\label{eq:stateprob}
P(\usetstar, V, \gamma |N, p, T, \etaset )= \frac{ V e^{-{\cal H}_{\gamma}( \usetstar , V)}}{\tilde{\cal Z}(N,p,T,\etaset)}  
 \end{equation}

Consider now a MC move (particle translation, association swaps, or  volume
move) which leaves the phase label unchanged. Let 

\begin{equation}
P_A \equiv P(\usetstar,V) \prod_{i=1}^{N-1} d\vec{u}_i dV
\end{equation}
be the probability that the system is found in the region of configuration space of volume
$\prod_i d\vec{u}_i dV$ around $\usetstar,V$. Similarly let 

\begin{equation}
P_B \equiv P(\usetstarprime, V^{\prime}) \prod_{i=1}^{N-1} d\vec{u}^{\prime}_i dV^{\prime}
\end{equation}
represent the probability associated with the range of configuration
space reached by implementing a change
$(\usetstar,V)\to(\usetstarprime,V^\prime)$. Detailed balance requires

\begin{equation}
\frac
{p_a( A \rightarrow B)}
{p_a( B \rightarrow A)}
=\frac{P_B}{P_A} 
\end{equation}
where

\begin{equation}
\frac{P_B}{P_A} =\frac{
V^{\prime} \exp[-{\cal H}_{\gamma} ( \usetstarprime , V^{\prime})]
\prod_{i=1}^{N-1} d\vec{u}^{\prime}_i dV^{\prime}
}
{
V \exp [-{\cal H}_{\gamma}( \usetstar ,  V)] 
\prod_{i=1}^{N-1} d\vec{u}_i dV
}\:.
\label{eq:db}
\end{equation}

Now for a volume update that {\em increments} the current value, $dV=dV^{\prime}$, while
\begin{equation}
\frac{V^{\prime} \prod_{i=1}^{N-1} d\vec{u}^{\prime}_i}{V\prod_{i=1}^{N-1} d\vec{u}_i}=
\left[ \frac{V^{\prime}}{V}\right]^N
\end{equation}

Thus
\[
\frac
{p_a( A \rightarrow B)}
{p_a( B \rightarrow A)}
=\frac{
\exp[-{\cal H}_{\gamma}( \usetstarprime ,  V^{\prime}) 
+N\ln V^{\prime}]}
{\exp[-{\cal H}_{\gamma}( \usetstar ,  V) +N\ln V]} 
\]
from which the  acceptance probability follows as

\begin{eqnarray}
\label{eq:moveacc}
\lefteqn{p_a (\usetstar, V \rightarrow \usetstarprime, V^{\prime}\mid \gamma)=}\nonumber\\
& & {\rm min} \left\{1,\exp\left[-\Delta {\cal H}_\gamma+  N\ln(V^{\prime}/V) \right]\right\}
\end{eqnarray}

with 
\begin{equation}
\Delta{\cal H}_\gamma={\cal H}_{\gamma}( \usetstarprime , V^{\prime}) -{\cal H}_{\gamma}( \usetstar, V)  \:.
\end{equation}

We note that the acceptance formula eq.~\ref{eq:moveacc} is actually no
different to that pertaining to a fully unconstrained constant-$NpT$
ensemble \cite{FRENKELSMIT}. Indeed a little thought reveals that the act of
fixing one particle is simply equivalent to viewing the unconstrained
system from the reference frame of that particle. Such a change of
reference frame has no consequences for acceptance criteria.
 
Consider next the inter-phase switch. This replaces one set of
representative vectors, $\Rsetgamma$  say, by the other,
$\Rsetgammaprime$, while at the same time scaling the system volume and
particle displacements by a constant factor chosen such that a
`typical' volume $\hat{V}_\gamma$ of phase $\gamma$ is matched to a
`typical' volume $\hat{V}_{\gamma^\prime}$ of phase $\gamma^\prime$.
For a volume update which {\em scales} the current value, one has $dV^\prime/dV=
V^\prime/V$, and referring back to the detailed balance relation
eq.~\ref{eq:db} one finds

\begin{equation}
\frac{V^{\prime} \prod_{i=1}^{N-1} d\vec{u}^{\prime}_i dV^\prime}{V\prod_{i=1}^{N-1} d\vec{u}_idV}=
\left[ \frac{V^{\prime}}{V}\right]^{N+1}\:.
\end{equation}
The switch acceptance probability follows as

\begin{eqnarray} 
\lefteqn{p_a (\gamma, \usetstar, V \rightarrow \gamma^{\prime}, \usetstarprime,  V^{\prime}) =}\nonumber \\
& &  \mbox{min} \left\{1,\exp\left[ \Delta {\cal H}_{\gamma^\prime\gamma} + (N+1) \ln(\hat{V}_{\gamma^{\prime}}/\hat{V}_{\gamma})\right]\right\}
\end{eqnarray}

where 

\begin{equation}
\Delta {\cal H}_{\gamma\gamma^\prime}= {\cal H}_{\gamma^{\prime}}( \usetstarprime ,  V^{\prime})-{\cal H}_{\gamma}(\usetstar , V)  \:.
\end{equation}
Note that a different formulation of the switch acceptance
probability applies when the displacement vectors are held constant
during the switch operation \cite{WILDING00,ERRINGTON04,NOTE0}.


\begin{thebibliography}{50}
\bibitem{OTHERAPPROACHES} This is evidenced by the persistence of
attempts to develop `other' approaches.
\bibitem{TEMPERINGSWAPS} The actual implementation swaps replicas
between the overlapping windows. These swaps are accepted with probability
unity whenever two adjacent replicas are in the overlap region, and are
rejected otherwise. In the central region, which has two exactly
overlapping windows --one for liquid and one for solid-- trail swaps are
accepted with unit probability, yielding a `density of states' which is
the simple average of those for the solid and liquid branches.
\bibitem{MASTNYCONTACT} At the time of writing, the authors of
ref.~\protect\cite{MASTNY05} are investigating whether repeated transitions
between the two phases occur in their method: J.J. de Pablo, personal
communication.
\bibitem{units} $p,V$ and $\Phi$ are expressed in units of
$kT/\sigma^3$, $\sigma^3$ and $kT$ respectively, where for a given interparticle potential $\sigma$ sets the
particle length scale, cf.~eq.~\protect\ref{eq:lj}.
\bibitem{glass} Unless the fluid phase configuration is `close' to the
representative one, the switch may create defective or glassy (not
crystalline) configurations.
\bibitem{NOTE0} In this work, the scaling at the switch is applied to
the box size, the representative configuration vectors and the displacement vectors.
Previous work \cite{WILDING00,ERRINGTON04} chose to keep the
displacement vectors constant at the switch. Both methods are
legitimate provided that the consequences for the switch acceptance
probability (see appendix.~\protect\ref{sec:append1}) are accommodated.
\bibitem{NOTE1} To ensure that no two sites of $\RsetF$ are atypically
close to one another (which might otherwise impede access to the
gateway states), we `anneal' this configuration before adopting it.
This entails quenching the system (at constant volume), while
suppressing crystallization by constraining particles to remain within
a maximum displacement $0.3\sigma$ from their starting point.
\bibitem{ERRORSNOTE} Coexistence parameters are determined inherently
more accurately by the PSMC method than the number and energy densities
of the respective phases (cf.~eq.~\protect\ref{eq:error}). 
One could therefore reduce (at relatively little cost) the uncertainties
associated with the densities simply by performing additional single
phase simulations at the stated coexistence parameters.
\end{thebibliography}
\end{document}